\documentclass[english,oneside]{elsart}
\usepackage{ae,aecompl}
\usepackage[T1]{fontenc}
\usepackage[latin1]{inputenc}
\usepackage{color}
\usepackage{graphicx}
\usepackage{amssymb}
\usepackage{natbib}

\makeatletter



\usepackage{babel}
\makeatother
\begin{document}
\begin{frontmatter}

\title{Improving Raman velocimetry of laser-cooled cesium atoms by spin-polarization}

\author{Julien Chab\'{e},}
\author{Hans Lignier\thanksref{HL},}
\author{Pascal Szriftgiser,}
\author{Jean Claude Garreau}

\thanks[HL]{Present address: Dipartimento di Fisica ``E. Fermi\char`\",
Universit\`a di Pisa, Pisa, Italy}

\address{Laboratoire de Physique des Lasers, Atomes et Mol\'{e}cules, UMR
CNRS 8523, Centre d'Études et de Recherches Laser et Applications,
Université des Sciences et Technologies de Lille, F-59655 Villeneuve
d'Ascq Cedex, France.\thanksref{WEB} }

\thanks[WEB]{http://www.phlam.univ-lille1.fr/atfr/cq}

\begin{abstract}
We study the peformances of Raman velocimetry applied to laser-cooled, spin-polarized,
cesium atoms. Atoms are optically pumped into the $F=4$, $m_{4}=0$ ground-state 
Zeeman sublevel, which is insensitive to magnetic perturbations. High resolution Raman 
stimulated spectroscopy is shown to produce Fourier-limited lines, allowing, in realistic
experimental conditions, atomic velocity selection to one-fiftieth of a recoil velocity. 
\end{abstract}
\end{frontmatter}

\section{Introduction}

Raman stimulated spectroscopy has been one of the most fertile techniques
used for manipulating laser-cooled atoms. It has been used for atomic
velocity selection \cite{Chu:RamanVSel:PRL91}, sub-recoil laser
cooling \cite{Chu:RamanCooling:PRL92,Phillips:RamanCooling:PRA2004}
and quantum state preparation and detection 
\cite{Chu:SideBand:PRL98,Salomon:SideBand:PRL99},
with applications in as different fields as quantum chaos 
\cite{AP:Bicolor:PRL00,Raizen:FluctDecChAssistTunnel:PRL02},
quantum dynamics in optical lattices 
\cite{Salomon:BlochOsc:PRL96,Raizen:LandauZennerWS:PRL96,Raizen:WSOptPot:PRL96},
quantum information processing \cite{Meschede:RamanSpec:APB04},
and high-precision metrology of fundamental constants 
\cite{Chu:MeasHoverMCs:PRL93,Biraben:MeshsurMRb:PRL04,Biraben:FineStrucCstBlochOsc:PRL06}.
Being a \emph{stimulated} two-photon transition between two ground-state
hyperfine sublevels, the width of the Raman line is, in principle,
limited only by the duration of interaction between the atom and the
light (the Fourier limit), as no natural widths are involved in the
process. These sharp transitions can thus be used to select a very
thin velocity class\emph{.} A laser radiation of wavelength $\lambda_{L}$
produces optical potentials with a typical well width of the order of
$\lambda_{L}/2$ (e.g. in a standing wave), whereas the so-called
\emph{recoil velocity} $v_{r}=\hbar k_{L}/M$ (with $k_{L}=2\pi/\lambda_{L}$
and $M$ the mass of the atom) corresponds to a de Broglie wavelength
$\lambda_{dB}=\lambda_{L}$. One thus sees that the deep quantum regime
$\lambda_{dB}\gg\lambda_{L}/2$ implies very sharp, ``sub-recoil'',
velocity distributions with $\langle v^{2}\rangle^{1/2}\ll v_{r}$,
which cannot be obtained in a simple magneto-optical trap.

For the alkaline atoms, the ground state presents hyperfine Zeeman
sublevels, so that the Zeeman effect and light shifts inhomogeneously
broaden the Raman transition. In order to obtain sharper lines one
must pump the atoms into a particular sublevel, avoiding inhomogeneous
broadening. In this respect, the $m_{F}=0$ Zeeman sublevel
\footnote{Throughout this paper, except otherwise indicated, we use the following
convention to note Zeeman sublevels: ground state sublevels are noted
$F,m_{F}$ and excited-state sublevels are noted $F^{\prime},m_{F^{\prime}}$.%
} is particularly interesting, as this level is not
affected by the first order Zeeman effect, and the second order effect
is negligible for the low magnetic fields we are considering here
\footnote{The order of magnitude of the {\em broadening} induced by
the second order Zeeman effect in our setup is 1 mHz.}.

Atom spin polarization has been used in many recent experiments in
various fields. In metrology, it was used in building frequency standards
\cite{Chu:CsFreqStandard:PRL93}, measuring the ratio $h/M$ in cesium
\cite{Chu:MeasHoverMCs:PRL93} and rubidium \cite{Biraben:MeshsurMRb:PRL04},
and in a recent determination of the fine structure constant \cite{Biraben:FineStrucCstBlochOsc:PRL06}.
It has also played an important role in optical dipole traps and sideband
cooling \cite{Wieman:PolarizedDipoleTrap:PRL99,Salomon:PolRaman:EL99,Weiss:SideBandCooling:PRL00}.
None of these works, however, concentrate in the polarization process
itself. The aim of the present work is thus to study in greater detail
the polarization process in the context of Raman velocimetry of cold
atoms. We describe a setup allowing optical pumping of laser-cooled
cesium atoms into the $F=4$, $m_{4}=0$ ground-state hyperfine sublevel,
concentrating on the improvement of the sensitivity of the Raman velocimetry
(RV) technique. This allows us, moreover, to measure and minimize the heating
induced in the atoms by the polarization process itself. 
We achieve a degree of polarization of $\sim$75\%
of the atoms in the $m_{4}=0$ sublevel with an increase of the \emph{rms}
velocity limited to 20\%. The observed Raman transition full width
at half maximum (FWHM) is 160 Hz, which corresponds to a velocity
resolution of $v_{r}/50$ (or 70 $\mu$m/s, the best reported velocity
resolution to our knowledge), to be compared to the $v_{r}/2$ resolution
of the compensated-magnetic-field line we observed in the same
experimental setup \cite{AP:RamanSpectro:PRA01} with unpolarized atoms.
We show that the Raman line resolution (in the velocity-independent
case) is Fourier-limited, i.e. the width of the line is comparable
to the inverse of the duration of the Raman pulse. In this sense,
we can say that we approached the RV ultimate limit.

\section{Experimental setup 
\label{sec:ExperimentalSetup}}

In this section, we briefly discuss basic aspects of atom spin polarization.
We consider specifically the case of the cesium atom, although most
of our conclusions can be easily extended to other alkalis. Atomic
spin polarization is performed in the presence of a \emph{bias magnetic
field}, which defines a fixed quantization axis for all atoms.

\begin{figure}
\begin{center}
\includegraphics[clip,width=10cm]{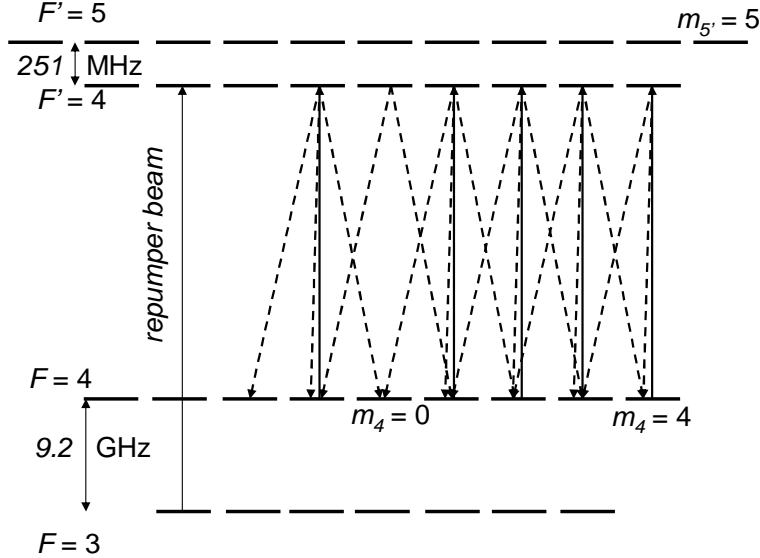}
\end{center}

\caption{ 
\label{fig:PumpingScheme}Pumping atoms with a $\pi$-polarized
radiation on a $F\rightarrow F$ transition polarizes the atoms in
the magnetic field-insensitive $m_{4}=0$ Zeeman sublevel. Laser-induced
transitions are represented by solid arrows, spontaneous-emission
transitions by dashed arrows. In order to keep the figure readable,
we did not represented all possible transitions. }
\end{figure}

\begin{figure}
\begin{center}
\includegraphics[width=10cm]{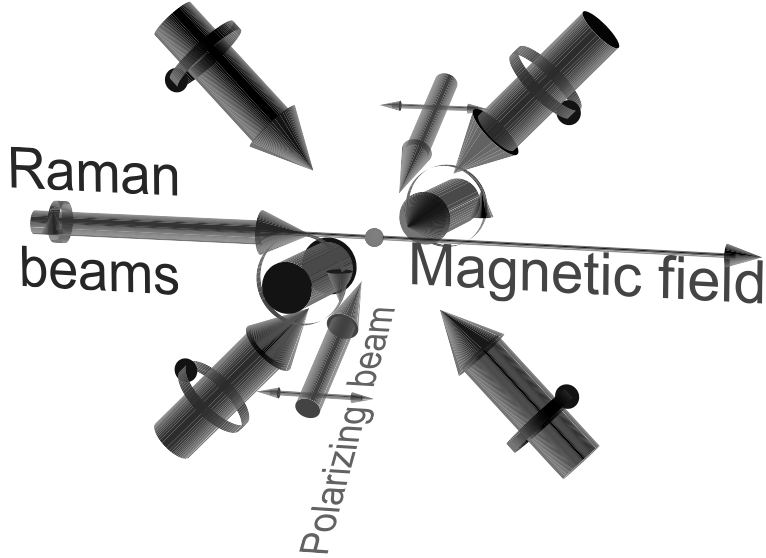}
\end{center}

\caption{
\label{fig:Setup}Experimental setup. The atomic cloud is at the
center. Three back-reflected beams (thick arrows) provide magneto-optical
trapping and cooling, two $\sigma^{+}$ Raman beams (thin arrows)
(displayed here in the copropagating configuration) induce Raman transitions.
The back-reflected polarizing beam (thinner arrows) is linearly polarized
along the bias magnetic field. The Raman beams are horizontal and
aligned with the bias magnetic field, while the PB is orthogonal to
it, making a 45$^{\circ}$ angle with respect to the horizontal.}
\end{figure}

The polarization technique we use, schematically presented in 
Fig.~\ref{fig:PumpingScheme},
is based on the fact that the Clebsch-Gordan coefficient coupling
the ground state sublevel $F,m_{F}=0$ to the excited sublevel 
$F^{\prime}=F,m_{F^{\prime}}=0$
vanishes. This means, for cesium, that if one optically pumps the
atom with \emph{linearly polarized light} on the $F=4\rightarrow F^{\prime}=4$
transition, the $m_{4}=0$ level is a dark state, in which atoms are
trapped. Because they are in a dark state, they are not submitted
to spontaneous emission heating due to fluorescence cycles. The transition
is however not closed, as the atoms can spontaneously decay from the
excited $F^{\prime}=4$ level to the $F=3$ level, so that one must
use a repumper laser beam coupling the levels $F=3$ and $F^{\prime}=4$,
which brings those atoms back into the polarizing cycle.

Our experimental setup (fig. \ref{fig:Setup}) consists of a standard
magneto-optical trap (MOT), a polarizing beam, a repumper beam and
a stimulated Raman spectroscopy setup. This setup can be used in
two different configurations. If the Raman beams are copropagating,
one can measure individual Zeeman sublevel populations, as the
dependence in the atomic velocity due to Doppler effect
cancels out in such case. If the beams are counterpropagating, 
the Doppler effect makes the transition probability dependent on the velocity, 
and the setup can be used to measure velocity class populations, and 
thus to reconstruct velocity distributions.
Our laser-cooling sequence includes a 25 ms-long Sisyphus-molasses phase with a large
detuning (-6 $\Gamma$) and small intensity (1\% of the saturation
intensity) which allows us to achieve a final temperature around 3 $\mu$K,
close to the Sisyphus limit temperature \cite{Salomon:SisyphusLimitT:EL90}.
The polarizing beam (PB), whose polarization axis is parallel to the bias magnetic
field, is extracted from the same diode laser that produces the MOT
beams. An independent acousto-optical modulator controls its frequency.
After interacting with the atom cloud, the PB is reflected back on
the cloud by a mirror, preventing the atoms to be pushed out of the
axis of the setup by the radiation pressure. The typical incident
power on the atom cloud is 2.5 $\mu$W.
More details on our experimental setup can be found in previous publications
\cite{AP:DiodeMod:EPJD99,AP:RamanSpectro:PRA01}.

Atoms issued from the MOT setup are mostly in the $F=4$ ground-state
hyperfine level. The Raman-resonant atoms are transferred by a Raman
pulse to the $F=3$ level and the atoms remaining in the $F=4$ level
are pushed out of the interaction region by a resonant pushing-beam
pulse. Atoms in the $F=3$ level are then optically repumped to the
$F=4$ level where they are excited by (frequency-modulated) resonant
light and their fluorescence is optically detected with a lock-in
amplifier.

The magnetic field fluctuations are reduced in our setup by an
active compensation scheme: small coils with the axis oriented along
the three orthogonal directions and located at opposite corners of
the cesium-vapor cell measure the magnetic field fluctuations, which
are electronically interpolated to deduce a value at the \emph{center}
of the cell. This error signal is used to generate currents sent through
3 mutually orthogonal Helmholtz coil pairs that generate a compensating
field. A constant bias current can also be applied to
the compensating coils in order to correct the DC component of the magnetic
field%
\footnote{In order to adjust the bias current, we minimize the
width of the Raman line in the copropagating configuration.} 
(we verified that this component remains stable to
better than 1\% over periods of time of one hour). This scheme allows
thus a reduction of the magnetic field fluctuations even if one wants to
keep a constant, non zero, bias field.
We measured a residual magnetic field {\it rms} fluctuation of 300 
$\mu$G \cite{AP:RamanSpectro:PRA01}.

In the present work we use $\sigma^{+}$-polarized Raman beams 
\cite{Meschede:RamanSpec:APB04}, so
that a given Raman transition involves only two Zeeman sublevels 
$m_{F}\rightarrow m_{F}+1\rightarrow m_{F}$,
and the intensity of each line in the Raman spectrum measures the
individual sublevel populations.

\section{ 
\label{sec:Results}Polarizing the atoms}

\begin{figure}
\begin{center}
\includegraphics[clip,width=10cm]{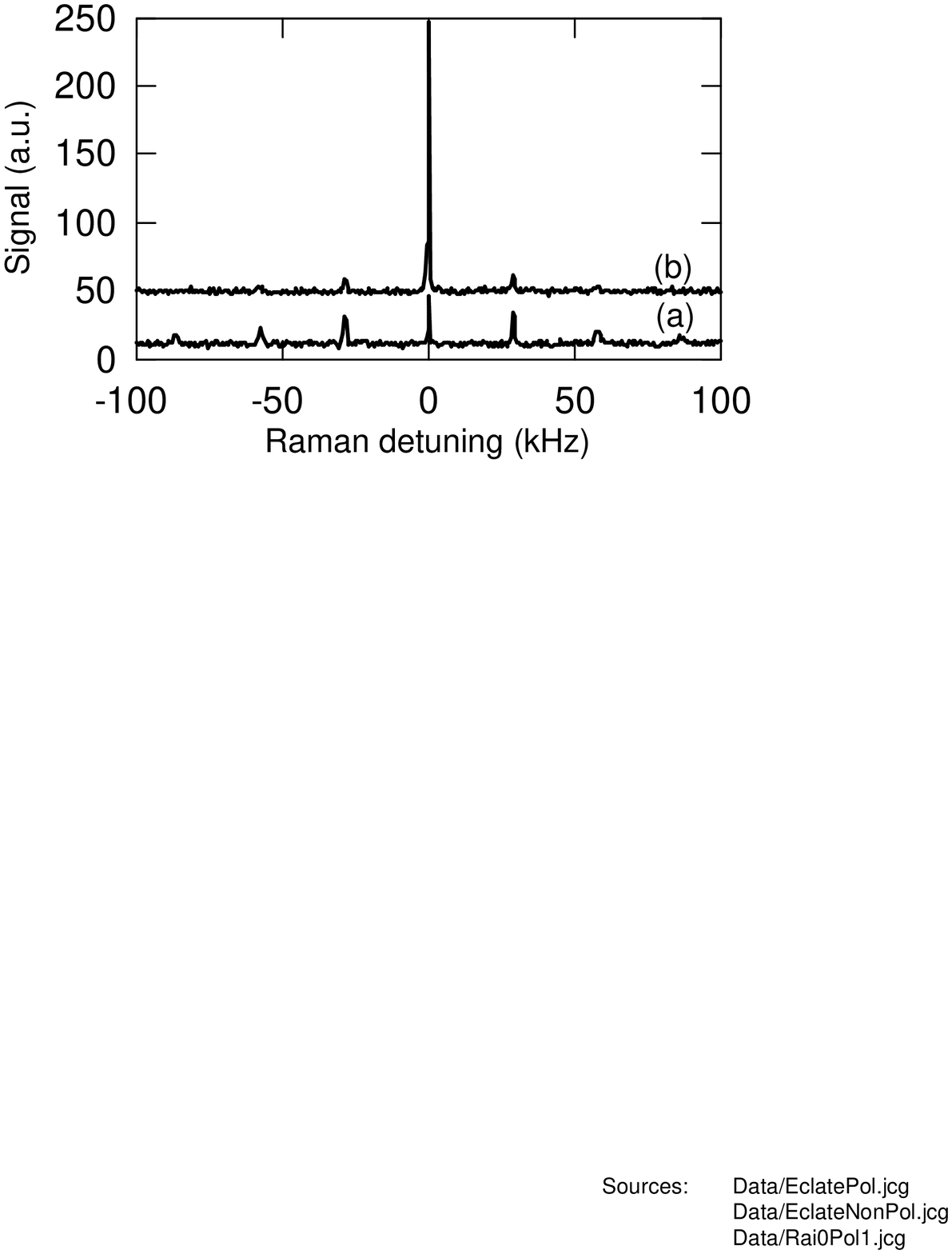}
\end{center}

\caption{ 
\label{fig:PolarizedCo}Copropagating-beam Raman spectra. (a) No
polarizing beam applied. (b) Polarizing beam applied (this plot was
vertically shifted in order to easy comparison); 75 \protect\% of the atoms
are in the $m_{4}=0$ Zeeman sublevel. The power of the PB is 2.5
$\mu$W and its detuning -0.5 $\Gamma$. }
\vspace{1cm}
\end{figure}

In order to measure the polarization, we perform Raman stimulated 
spectroscopy with \emph{copropagating} beams, for
which the transition is insensitive to the atomic velocity. The bias
magnetic field is adjusted so that Raman lines are clearly separated.
Fig.~\ref{fig:PolarizedCo} compares the Raman spectra obtained without
(a) and with (b) polarization. In presence of the PB, 75\% of the
atoms have been pumped into the $m_{4}=0$ sublevel. The fact that
not all the atoms are in the dark state can be attributed to experimental
imperfections in defining the polarization of the PB beam, due to
its transmission through the MOT cell walls at an angle that is not
exactly 90$^{\circ}$, and to a residual misalignment between the PB polarization
and the bias magnetic field, whose direction varies a little bit across
the atom cloud. Polarizations higher than 95\% were reported in the literature, 
\cite{Wieman:PolarizedDipoleTrap:PRL99,Steck:PhDThesis:01},
but optimizing our setup to such level would imply an overall redesigning,
which is not worth a gain of $\sim 20$\% in the atom number. By comparing
the total area in all lines in the two spectra, we deduce that about
20\% of the atoms were lost in the polarization process, which can
be attributed essentially to the atom cloud free fall due to gravity.

\begin{figure}
\begin{center}
\includegraphics[clip,width=10cm]{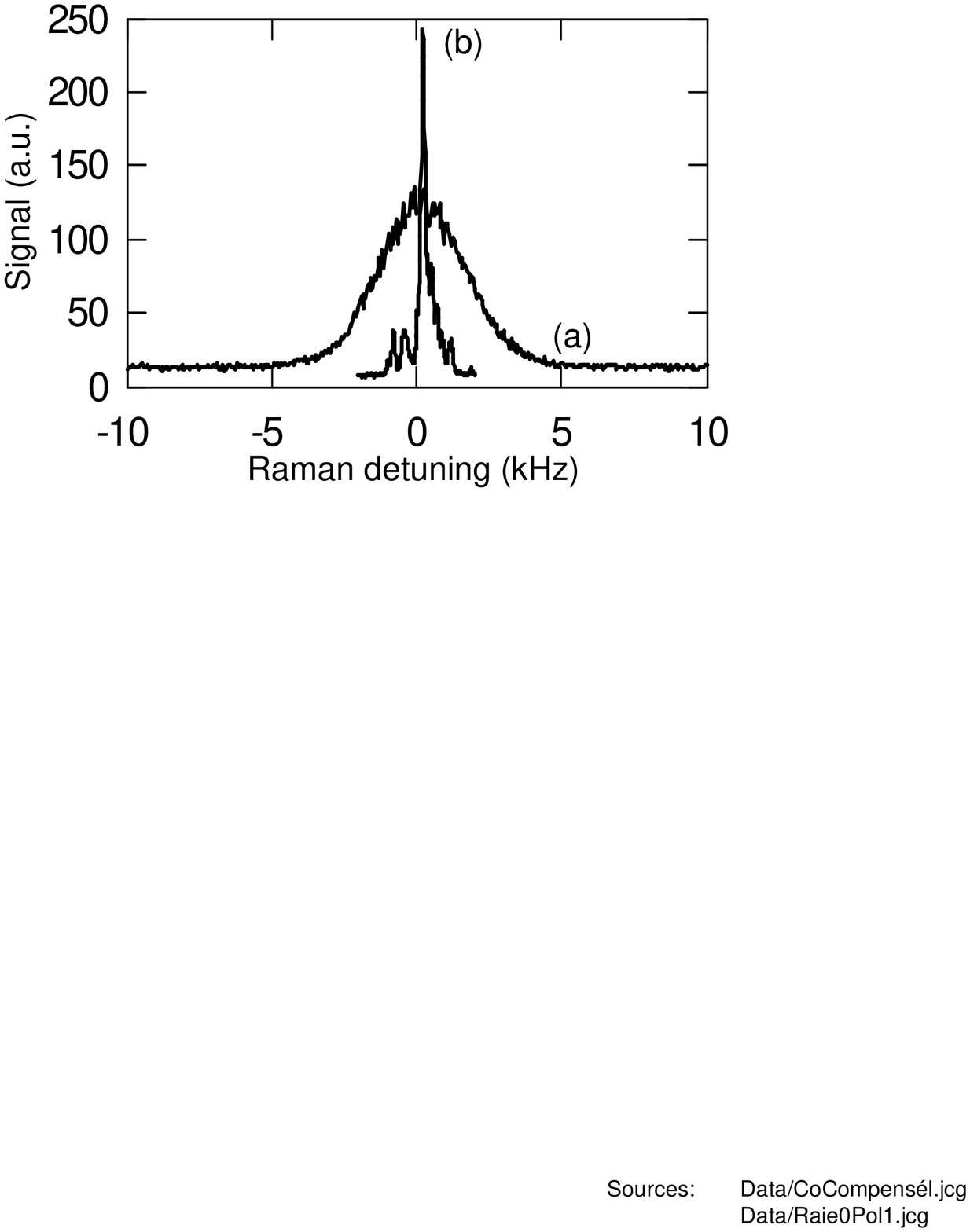}
\end{center}

\caption{
\label{fig:Comp}Comparison between copropagating-beam Raman spectra:
(a) the unpolarized line obtained with no bias field (3.5 kHz FWHM) and
the active compensation of the magnetic field on, 
and (b) the polarized $m_{4}=0$ line. The FWHM
of the polarized line is 160 Hz, which implies a velocity resolution
of $v_{r}/50$.}
\end{figure}

In Fig.~\ref{fig:Comp}(a) we show the line obtained with unpolarized
atoms, bias magnetic field tunned to zero, and the active compensation
of field fluctuations on. This line includes contributions from 
all Zeeman sublevels, and its width can be attributed to inhomogeneous
broadening of the various components due to light-shift
and residual magnetic-field fluctuations \cite{AP:RamanSpectro:PRA01}.
This line is compared to the single polarized $m_{4}=0$ line. 
The improvement factor in the FWHM is 22.
The measured FWHM of the polarized line is 160 Hz, which, when multiplied
by the Raman pulse duration (7 ms) gives 1.12: we are thus very close
to the Fourier limit. The structures in the pedestal of the polarized
line are due to weak $m_{4}=0\rightarrow m_{3}=\pm1$ and $m_{4}=\pm1\rightarrow m_{3}=0$
transitions due to the fact that the Raman beams wavevectors are not
perfectly aligned with the bias magnetic field. In
the counterpropagating case, the Doppler broadening prevents 
a direct measure of the resolution. However, as the perturbing
effects are identical as in the copropagating case, it is rather safe
to assume that the resolution is the same also in both cases.

The fluorescence cycles performed by the atom during the polarization
process inevitably induce spontaneous-emission heating. In order to
evaluated this heating effect, we compared the lines obtained
in the \emph{counterpropagating} Raman beam configuration with and without polarization.
The observed width of 160 Hz in the polarized case corresponds, when 
extrapolated to the counterpropagating configuration, to a velocity resolution
of $\sim0.02v_{r}$, to be compared to $\sim0.4v_{r}$ in the unpolarized
case. Table \ref{tab:Heating} displays the parameters used and the
observed linewidths. A back-reflected PB produces a minimum of heating
for a detuning of $-0.5\Gamma$, corresponding to the minimum temperature
of the Doppler cooling. We then observe an increase in the \emph{rms} velocity
of only 20\%, from 4.0 to 4.8 $v_{r}$. We can roughly
interpret this heating effect by calculating the number of fluorescence
cycles necessary to bring an atom from a Zeeman substate $m_{F}$
to the substate $m_{F}=0$ and averaging over both $m_{F}$ and over
the random direction of the spontaneously-emitted photons. We obtain
a value for the increase in the \emph{rms} velocity
of 1.1 $v_{r}$, which matches well with the experimental values shown
in Table \ref{tab:Heating}.

\begin{table}
\begin{center}
\begin{tabular}{c c c c c}
\hline
\mbox{$\Delta_{PB}$} & $\tau$ & $P$ & FWHM & $v_{rms}$ \\
($\Gamma$) & (ms) & ($\mu$W) & (kHz) & ($v_r$) \\
\hline
-1.0 & 0.53 & 2.41 & 98.6 & 5.2 \\
-0.5 & 0.32 & 2.34 & 91.3 & 4.8 \\
0.0 & 0.65 & 2.36 & 99.6 & 5.2 \\
No PB & -- & -- & 75.3 & 4.0 \\
\hline
\end{tabular}
\end{center}
\caption{
\label{tab:Heating}Heating induced by the polarization process. 
Parameters are the detuning $\Delta_{PB}$ and the power $P$
of the polarizing beam and $\tau$ the total duration of the polarization
process (the duration is chosen so that 50\protect\% of the atoms are 
pumped into the $m_{4}=0$ level).}
\end{table}

\begin{figure}
\begin{center}
\includegraphics[clip,width=10cm]{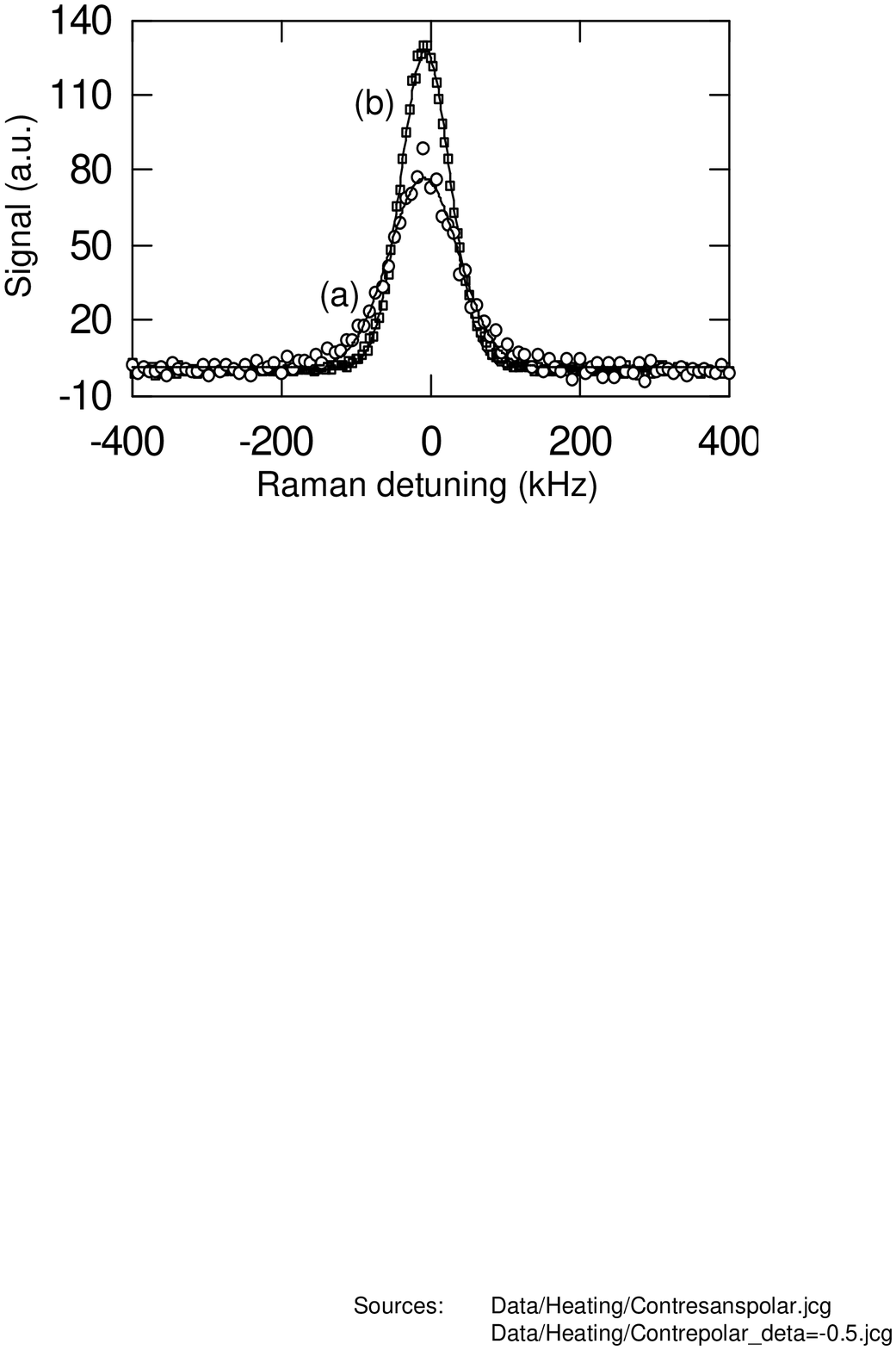}
\end{center}

\caption{
\label{fig:Counterprop}Counterpropagating-beam, velocity-sensitive
Raman spectra. (a) Spectrum obtained with atomic polarization on the
$m_{4}=0$ line (empty circles); (b) spectrum obtained with active
compensation of the magnetic field (empty squares). Both lines are
well fitted by Gaussians, from which one deduces \emph{rms} velocities
of, resp., 4.8 and 4.0 $v_{r}$, (4.6 and 3.2 $\mu$K). For comparison,
the recoil velocity corresponds to $8.27$ kHz of Raman detuning.}
\end{figure}

Fig.~\ref{fig:Counterprop} compares the counterpropagating-beam
spectra obtained with and without the application of the PB. They
are very well fitted by Gaussians, and can be considered as directly
proportional to the velocity distribution. The ratio of the surfaces
of the two distributions is $\sim$11\% which constitutes a rather
acceptable loss. We also observed a decrease of a factor 1.8 in the signal to
noise ratio in the polarized case. This is probably due to the fact that 
the number of atoms present in the
cloud has larger fluctuations because
of the polarization process itself, as compared to the fluctuations observed
in the absence of polarization. That is, the polarization both reduces
the number of atoms and increases its fluctuations, thus degrading
the signal to noise ratio more strongly than the observed loss
in the atom number.

\begin{figure}
\begin{center}
\includegraphics[clip,width=10cm]{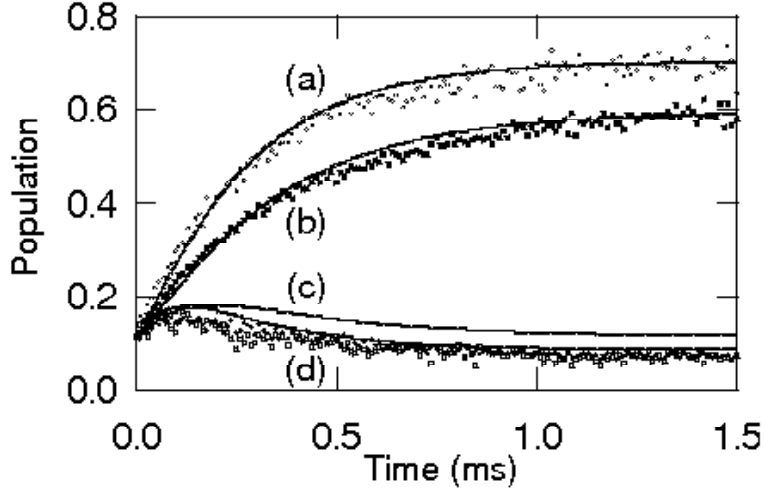}
\end{center}

\caption{
\label{fig:Dynamics}Dynamics of the polarization process. (a) Population
of the $m_{4}=0$ level with $\Delta_{PB}=-0.5\Gamma$, (solid lines
are fits of the experimental data by numerical simulation curves); (b) Population of the $m_{4}=0$
level with $\Delta_{PB}=0$; (c) Population of the $m_{4}=1$ level
with $\Delta_{PB}=0$; (d) Population of the $m_{4}=1$ level with
$\Delta_{PB}=-0.5\Gamma$. Parameters are $I_{PB}/I_{s}=0.019$, $I_{repumper}/I_{s}=0.023$,
$\alpha=0.013$ {[}cf. Eq.~(\ref{eq:contamination})].}
\end{figure}

Finally, we have modelized the dynamics of the polarization process,
and quantified the main stray effect, namely the polarization defect of the
PB. Fig.~\ref{fig:Dynamics} shows a comparison of the experimentally
observed level populations and the evolution predicted by a rate-equation
model described in appendix \ref{sec:Model}.
The simulation fits very well the experimental results, except for the population
of the $m_{4}=1$ level with $\Delta_{PB}=0$, where one observes
a systematic shift, probably due to the fact that these small values
are close to the limit detection level. We deduce a rather small depolarization
value of $\alpha=0.013$ [see Eq.~(\ref{eq:contamination})] , 
which shows that the polarization process is very sensitive
to this effect.

\section{Conclusion}

We evidenced the ability of the polarization technique to
produce very sharp Raman lines, allowing high-resolution Raman velocimetry
of laser-cooled atoms. Our results imply a velocity resolution of
70 $\mu$m/s, or $v_{r}/50$, which can be compared to $v_{r}/18$
reported in \cite{Salomon:BlochOsc:PRL96}, $v_{r}/17$ in \cite{Steck:PhDThesis:01}
(both with cesium) and 290 $\mu$m/s, or $v_{r}/100$, in \cite{Chu:RamanVSel:PRL91}
(for sodium). The present work thus corresponds to the best observed
\emph{velocity} resolution. Such a resolution implies a de Broglie
wavelength of $50\lambda_{L}$, which potentially generates coherent
atomic wavefunctions extending up to 100 wells of a standing wave.
The technique is thus highly useful in manipulating the external degrees
of freedom of atoms in the frame of experiments on quantum dynamics.

\begin{ack}
Laboratoire de Physique des Lasers, Atomes et Molécules (PhLAM) is
Unité Mixte de Recherche UMR 8523 du CNRS et de l'Université des Sciences
et Technologies de Lille. Centre d'Études et de Recherches Laser et
Applications (CERLA) is supported by Ministère de la Recherche, R\'{e}gion
Nord-Pas de Calais and Fonds Européen de Développement Économique
des Régions (FEDER). 
\end{ack}
\appendix

\section{ 
\label{sec:Model}Model}

In order to understand the dynamics of the polarizing process dynamics,
we performed, as in refs. \cite{Cerez:PolCesium:PRA87,Sagle:CsD2Line:JPBAMOP96},
numerical simulations based on a rate-equation approach taking into
account the effect of the PB and of the repumper. We use in the present Appendix
a slightly different notation: we identify a given sublevel by three
labels: $s=\{ g,e\}$ characterizing fine-structure state, $F=\{3,4,5\}$
for the hyperfine sublevel and $m_{F}=\{-F...F\}$ for the Zeeman
sublevel. The general form of these equations is then 
\begin{eqnarray}
\frac{dN_{s,F,m_{F}}}{dt} & = & \sum_{s_{1},F_{1},m_{1}}W_{s_{1},F_{1},m_{1}\rightarrow s,F,m_{F}}N_{s_{1},F_{1},m_{1}}-\nonumber \\
 &  & \sum_{s_{1},F_{1},m_{1}}W_{s,F,m_{F}\rightarrow s_{1},F_{1},m_{1}}N_{s,F,m}+\nonumber \\
 &  & \delta_{s,g}\sum_{s_{1},F_{1},m_{1}}\Gamma a_{e,F_{1},m_{1}\rightarrow g,F,m_{F}}N_{g,F_{1},m_{1}}-\nonumber \\
 &  & \delta_{s,e}\Gamma N_{e,F,m}.
\label{eq:RateEq}
\end{eqnarray}
 where $N_{s,F,m_{F}}$ is the population of the sublevel $\{ s,F,m_{F}\}$,
$W_{s,F,m_{F}\rightarrow s_{1},F_{1},m_{1}}$ is the stimulated transition
rate between levels $\{ s,F,m_{F}\}$ and $\{ s_{1},F_{1},m_{1}\}$,
and $a_{e,F_{1},m_{1}\rightarrow g,F,m_{F}}$ is the spontaneous emission
branching ratio connecting the sublevels $\{ e,F_{1},m_{1}\}$ and
$\{ g,F,m_{F}\}$. The absorption and stimulated emission rates $W_{s,F,m_{F}\rightarrow s_{1},F_{1},m_{1}}$
are related to the spontaneous emission rates $a_{e,F_{1},m_{1}\rightarrow g,F,m_{F}}$
by the following relation: 
\begin{eqnarray}
W_{s,F,m_{F}\rightarrow s_{1},F_{1},m_{1}} & = & \frac{3}{2}
\frac{\lambda_{L}^{3}}{\pi hc}\frac{I}{\Delta_{L}}
\chi_{F\rightarrow F_{1}}a_{e,F_{1},m_{1}\rightarrow g,F,m_{F}}
\Gamma\varepsilon_{m_{F}-m_{1}}^{2}
\label{eq:rate}
\end{eqnarray}
 where $\lambda_{L}=852$ nm is the laser wavelength, $I$ the laser
intensity expressed in W$\cdot$m$^{-2}$ and $\Delta_{L}\sim2\pi\times1$
MHz the laser linewidth, and 
\begin{equation}
\chi_{F\rightarrow F_{1}}\equiv\mu(\mu+1)
\frac{\Delta_{F\rightarrow F_{1}}^{2}+(\mu-1)^{2}}
{(\Delta_{F\rightarrow F_{1}}^{2}+\mu^{2}-1)^{2}+4\Delta_{F\rightarrow F_{1}}^{2}}
\end{equation}
 is the relative probability of exciting a neighboring transition
with $\mu=\Delta_{L}/\Gamma$ and $\Delta_{F\rightarrow F_{1}}=2(\omega_{F\rightarrow F_{1}}-\omega_{L})/\Gamma$
the position of the atomic linewidth from the laser line. The branching
ratios $a_{e,F_{1},m_{1}\rightarrow g,F,m_{F}}$ can be found in \cite{Cerez:PolCesium:PRA87}
\footnote{Applying Eq.~(\ref{eq:RateEq}) to all transitions produces a set
of 43 coupled equations. We noted however that the ratio between different
values of $\chi_{F\rightarrow F_{1}}$ can be as large as $10^{4}$;
some transitions can in practice be neglected, and one obtains very
good results with only 23 equations :-).%
}.

As we indicated in Sec. \ref{sec:Results}, the polarization of the
PB is contaminated by $\sigma^{+}$ and $\sigma^{-}$ components.
This is taken into account in our simulation by writing its polarization
as

\begin{equation}
\mathbf{\mathbf{\varepsilon}}=\frac{\mathbf{\epsilon_{0}}+
\alpha\mathbf{\epsilon}_{+}+\alpha\mathbf{\epsilon}_{-}}
{\sqrt{1+2\alpha^{2}}}
\label{eq:contamination}
\end{equation}
$\alpha$ being an adjustable parameter representing the depolarization
of the PB. The coupled rate equations are numerically solved using
a standard $4^{\mathrm{th}}$ order Runge-Kutta integration method
with the initial condition that all atoms are in the $F=4$ ground-state
level and that all of its sublevels are equally populated.


\end{document}